\documentclass[%
onecolumn,%
oneside,%
floats,%
aps,%
 prd,%
nobibnotes,%
nofootinbib,%
amsmath,%
amssymb,%
amsfonts,%
amscd,%
  superscriptaddress,%
eqsecnum%
]{revtex4}

\usepackage[utf8]{inputenc}
\usepackage[english]{babel} 
\usepackage{graphicx}
\usepackage{xcomment}
\usepackage{bm}
\usepackage{float}
\usepackage{wasysym}

\newcommand\apjs{Astrophys. J. Suppl.}
\newcommand\apjl{Astrophys. J. Letters}
\newcommand\apss{Astrophys. Space Sci.}

\newcommand\mnras{Mon. Not. of the Royal Astro. Soc.}
\newcommand\aap{Astron. \& Astrophys.}

\newcommand\sovast{Soviet Astronomy}

\begin{document}

\title{Dynamics of supernova bounce in laboratory}

\author{S.I.~Blinnikov}
\email{Sergei.Blinnikov@itep.ru}
\affiliation{NRC ``Kurchatov Institute'' -- ITEP, Moscow 117218,  Russia}
\affiliation{Dukhov Research Institute of Automatics (VNIIA), Moscow 127050,  Russia}
\affiliation{Kavli IPMU (WPI), Tokyo University, Kashiwa 277-8583, Japan}
\affiliation{Space Research Institute (IKI), RAS, Moscow 117997, Russia}

\author{R.I.~Ilkaev}
\email{ilkaev@vniief.ru}
\affiliation{Russian Federal Nuclear Center VNIIEF, Sarov, Nizhni Novgorod region 607188, Russia}

\author{M.A.~Mochalov}
\email{postmaster@ifv.vniief.ru}
\affiliation{Russian Federal Nuclear Center VNIIEF, Sarov, Nizhni Novgorod region 607188, Russia}

\author{A.L.~Mikhailov}
\email{staff@vniief.ru}
\affiliation{Russian Federal Nuclear Center VNIIEF, Sarov, Nizhni Novgorod region 607188, Russia}


\author{I.L.~Iosilevskiy}
\email{iosilevskiy@gmail.com}
\affiliation{Joint Institute for High Temperatures, Russian Academy of Sciences, Moscow 125412, Russia}

\author{A.V.~Yudin}
\email{yudin@itep.ru}
\affiliation{NRC ``Kurchatov Institute'' -- ITEP, Moscow 117218,  Russia}

\author{S.I.~Glazyrin}
\email{glazyrin@itep.ru}
\affiliation{Dukhov Research Institute of Automatics (VNIIA), Moscow 127050,  Russia}
\affiliation{NRC ``Kurchatov Institute'' -- ITEP, Moscow 117218,  Russia}
\affiliation{Sternberg Astronomical Institute, Moscow M.V.~Lomonosov State University, Moscow 119234, Russia}
\affiliation{National Research Nuclear University MEPhI (Moscow Engineering Physics Institute),  Moscow 115409, Russia}

\author{A.A.~Golubev}
\email{alexander.golubev@itep.ru}
\affiliation{NRC ``Kurchatov Institute'' -- ITEP, Moscow 117218,  Russia}
\affiliation{National Research Nuclear University MEPhI (Moscow Engineering Physics Institute),  Moscow 115409, Russia}


\author{V.K.~Gryaznov}
\email{grvk@ficp.ac.ru}
\affiliation{Institute of Problems of Chemical Physics, Russian Academy of Sciences, Chernogolovka, Moscow region  142432, Russia}

\author{S.V.~Fortova}
\email{sfortova@mail.ru}
\affiliation{Institute of Computer Aided Design, Russian Academy of Sciences, Moscow 123056, Russia
}

%
%

%
%
%
%

\begin{abstract}
We draw attention to recent high explosive (HE) experiments which
provide compression of macroscopic amount of matter to high, even record, values of pressure in
comparison with other HE experiments.
The observed bounce after the compression corresponds to processes in core-collapse
supernova explosions after neutrino trapping.
Conditions provided in the experiments resemble those in core-collapse supernovae,
permitting their use for laboratory astrophysics.
A unique feature of the experiments is compression at low entropy.
The values of specific entropy are close to those obtained in numerical simulations during the process
of collapse in supernova explosions, and much lower than those obtained at laser ignition facilities,
another type of high-compression experiment.
Both in supernovae and HE experiments the bounce
occurs at low entropy, so the HE experiments provide a new platform to realize some
supernova collapse effects in laboratory, especially to study hydrodynamics of collapsing flows and the
bounce.
Due to the good resolution of diagnostics in the compression of macroscopic amounts of material with
essential effects of nonideal plasma in EOS, and
observed development of 3D instabilities,  these experiments may serve as a useful benchmark for
astrophysical hydrodynamic codes.
\end{abstract}

\maketitle

\section{Introduction}
\label{sec:introduction}

Astrophysics gathers data mainly from observations with no hope for full-scale
experiments under laboratory conditions~\cite{1998PhT....51h..26H, Drake, Fortov2016b, Fortov2016}, especially on supernovae -- the most energetic events in the universe.
Nevertheless, recent progress in high energy density laboratory experiments~\cite{1998PhT....51h..26H, Drake, Fortov2016b, Fortov2016} allows us to simulate to some extent the conditions in astrophysics.
The main sites of such experiments are cumulative high explosive (HE) generators~\cite{1998PhT....51h..26H, Drake, Fortov2016b, Fortov2016}, lasers~\cite{Henderson2003, remington:15}, pulsed high-current facilities (Z-machine)~\cite{1998PhT....51h..26H, Drake, Fortov2016b, Zexperiments, 2006RvMP...78..755R}.
Here we consider the HE driver shock-wave generator employed at explosive facilities.
The values of pressure most recently obtained there are at record highs for this type of experimental
facility: pressure ${P \sim 100}$~Mbar is reached in hydrogen isotope deuterium~\cite{mochalov:18} as we report here. 
Though it is significantly lower than the pressure inside exploding stars,
these experiments provide compression of a macroscopic amount of matter at
low entropy, leading to a shock wave bounce that corresponds to processes in core--collapse
supernova explosions.
Such a similarity appears due to the growth of the stiffness in the equation of state
due to strong quantum degeneracy effects for the free-electron component of
strongly nonideal deuterium plasma, an effect that is emphasized by low-entropy conditions.
Thus these experiments provide a new platform to realize some supernova collapse effects in the laboratory.

Supernova explosions represent one of the most energetic and exciting objects in the universe~\cite{reviewSN, Drake, Fortov2016b},  which is why researchers are so interested in studying them.
The records of those events are found in ancient chronicles during the millennia of written history, but the real scientific study of supernovae began only in 20th century.
During the past few decades a lot of theoretical models have been proposed~\cite{reviewSN},
none of which could describe all stages of this complex and extreme phenomenon.

At the same time, the general picture of the explosion is quite well established~\cite{reviewSN}. 
For initial stellar masses $\sim (10 \, - \, 25) M_\odot$, the explosion occurs due to core collapse.
In a stationary star the internal pressure is exactly compensated by gravity.
After exhaustion of nuclear fuel in the core of a star, the equation of state changes,
leading to the decrease of the adiabatic exponent when the internal pressure is unable to resist
the growth of the gravitational force in the core.
This eventually leads to unimpeded compression or collapse.
Unless another equation of state change happens, the process would go to infinity.
In reality, at high density due to degeneracy and nonideality of plasma, the
stiffness of the equation of state rises dramatically, which
leads to the bounce of infalling matter and formation of the outgoing shock, which
finally may produce the supernova explosion.
This process is complicated by other physical effects, the most important of which is the
generation of huge neutrino flux (see Sec.~\ref{stCollSimul}).

Today there is no hope to approach in the laboratory the ultraextreme conditions in plasma
that appear in supernova explosions~\cite{1998PhT....51h..26H, Drake, Fortov2016b, Fortov2016}, but we try to
qualitatively reproduce the hydrodynamical phenomena occurring in collapsing material: the
effects of degeneracy, nonideal plasma and bounce. A similar physics appears in
explosive experiments~\cite{Ashcroft1989,Ashcroft2002}
with
deuterium plasma.
Another benefit of these explosive experiments is in significantly lower values of specific
entropy compared to laser. This feature is important for comparison with core-collapsing supernovae,
where the values of entropy are also low.

The unique high energy laboratory experiments were specifically designed to reach high levels of
pressure and density.  
These conditions are reached as a result of a special assembly: a multilayer system that maintains a quasi-isentropic regime of
compression with low levels of entropy generation.
The achievements of recent years in symmetry control of explosive compression allow us
to reach high values of pressure. At the final stage the deuterium plasma has pressure
$P=114 \pm 20$~Mbar, and this is the record for shock-wave experiments with high explosives.
Generation of high-pressure, high-temperature  matter by multiple reverberating shock waves
is well known~\cite{Nellis1999,Ternovoi1999,Ternovoi2002} but is restricted by the pressure level
$\sim 1-5$~Mbar.

This paper has the following structure.
Section~\ref{experiments} discusses our HE experiments: the device and experimental results.
Section~\ref{thermodynamics} presents the equations of state that describe matter under extreme
conditions in explosion experiments (but not in supernova (SN) explosions).
Section~\ref{stCollSimul} presents our simulations of stellar collapse and describes the basic similarity between HE experiments and the core-collapse physics.
In Sec.~\ref{sec:summary} we summarize  our results and in Sec.~\ref{sec:conclusions} we
suggest a path for future work.

\section{High explosive experiments}
\label{experiments}

\begin{figure}[H]
\centering
\includegraphics[width=0.42\textwidth]{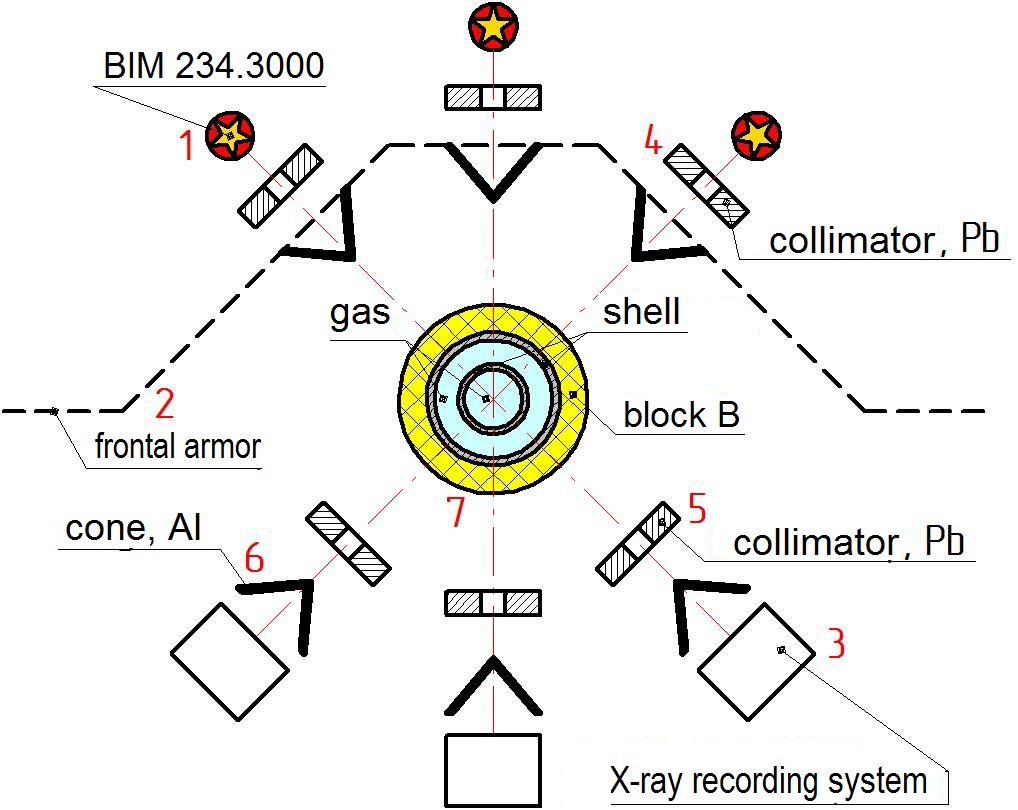}
\caption{Schematics of experiment. 1,  sources of x-ray radiation (betatrons);
2, shielding; 3, registrators; 4-5, collimators (Pb); 6, cone (Al);
7, target experimental device (working gas, shells, high explosive).
This design is for a two-cascade device with the central block shown in Fig.~\ref{fig:S2}.
In the case of a single-cascade device the central block is simpler, cf. Fig.~\ref{fig:2sphereXray}.
}
\label{fig:SphereMochalov}
\end{figure}

A high explosive experiment that includes high pressure studies of matter needs a
specially designed geometry. 
The target is a gas,
surrounded by a shell-pusher, that is accelerated by high-explosives outside the
shell. The installation is carefully adjusted with used explosive
intensity in order to reach the effective compression of the target. Also a number of diagnostics
are presented. The whole construction is schematically depicted in Fig.~\ref{fig:SphereMochalov}.

Using this construction we can observe various
stages of plasma dynamics in the x-ray images like the one shown in the left panel of
Fig.~\ref{fig:2sphereXray}.

\subsection{Single-cascade devices}

\begin{figure}[H]
\centering
\includegraphics[width=0.33\textwidth]{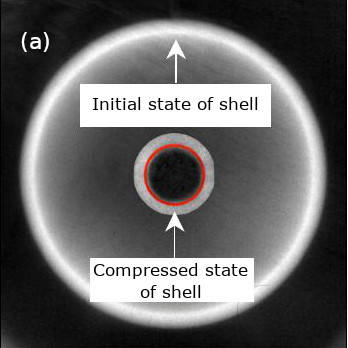}
\includegraphics[width=0.5\textwidth]{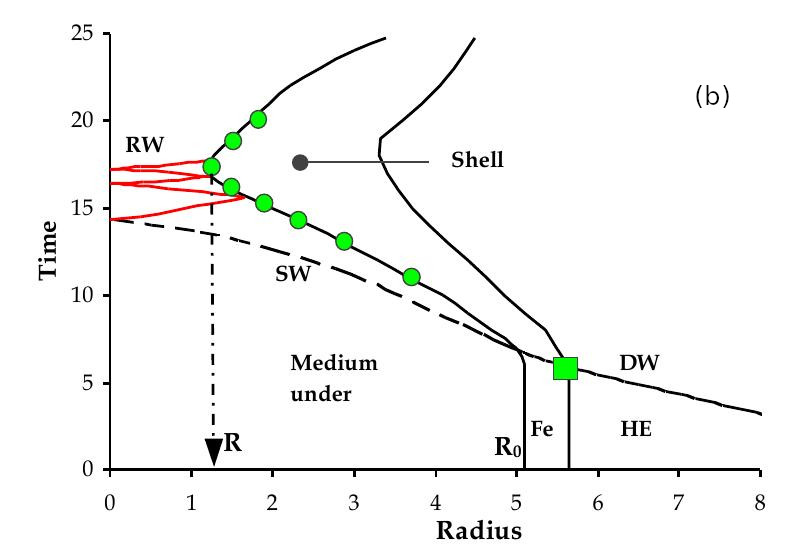}
\caption{(a) x-Ray images of spherical device with a single-cascade scheme~\cite{mochalov:10}. 
The image shows both the initial state and the  maximum compression state.
(b) Schematics of the single-cascade device dynamics.
Circles (green online) show the positions of internal shell boundary measured in the experiment.
The square dot shows the shell position measured by electrocontact technique.
DW, detonation wave; SW, shock wave; RW, reveberating wave; HE, high explosive.}
\label{fig:2sphereXray}
\end{figure}

The simplest experimental device (``a single-cascade scheme'') for studying
the compressibility of dense plasma is
a gas-filled metal shell of spherical shape surrounded by
a block of a HE.
Schematically, the process of matter compression
in such a construction is shown in Fig.~\ref{fig:2sphereXray} in the radius-time
diagram.
After the end of explosives detonation, a shock wave (SW) is formed in the metallic shell. Further, the SW enters the investigated substance.
After its reflection from the center of the system and from the inner boundary of the moving shell
a system of weak shock waves is formed in the region of the investigated substance (red lines) that compress and heat it up.
Additional compression is provided by the shell smoothly
converging to the center. Thus, the circulation of shock waves and smooth compression by the shell transform shock-wave compression into quasi-isentropic compression.
In the final result, due to the growth of pressure inside the material being
studied, the shell stops (at radius $R$ in Fig.~\ref{fig:2sphereXray}) and then bounces back.
Smaller is the jump in entropy in the first and subsequent shock waves, closer the
compression process approaches the isentropic one.
In such systems,
conditions for a longer retention of the matter at  high pressure are more favorable
as compared with loading by a single shock wave.
With this loading method, various isentropes are achieved,
with parameters depending on the mass of explosives and the geometry of the experimental devices.
Those parameters can be changed in a wide range.
Such a
single-cascade construction allows us to compress  hydrogen  to the density about
2~g/cm$^3$ by pressure $P\approx 1300 $~GPa~\cite{grigor:72, grigor:75}.
A similar construction was used also in Ref.~\cite{mochalov:10}, where the deuterium plasma was
compressed to the density of 4~g/cm$^3$ at pressure of 1800~GPa.

The physical picture developing in those experiments is rather simple.
First, we have a fluid of neutral molecular hydrogen that is dissociated and ionized
by the first and second shocks to almost ideal plasma.
This stage is followed by the fast growth of Coulomb corrections in EOS of
the compressed hydrogen (Coulomb nonideality) and electron degeneracy
in the second, third, etc., shocks~\cite{1998PhT....51h..26H, Drake, Fortov2016b, Fortov2016}.
Finally, a transition to strongly degenerate nonideal plasma occurs.
It should be noted that the degree of the Coulomb corrections is
controlled by a special dimensionless  parameter of
nonideality~\cite{1966JChPh..45.2102B,1969PhLA...28..706V,1980PhRvA..21.2087S,1989ASPRv...7..311Y,2010JPhA...43g5501G},
\begin{equation}
\Gamma = \frac{(Z e)^2}{r_s k T}  \approx \frac{e^2 n_e^{1/3}}{kT} ,
\label{nonIdGamma}
\end{equation}
where $Z=1$ for hydrogen and $r_s$ is the Wigner–Seitz radius, defined by
\begin{equation}
r_s =  \left( \frac{3}{4\pi n_e} \right)^{1/3} .
\label{WigSei}
\end{equation}
The parameter
$\Gamma$ is equal to the ratio of the Coulomb interaction
energy to the average kinetic energy of charged particles.
The pressure correction
due to electron degeneracy is controlled by another dimensionless
parameter,
\begin{equation}
\xi=n_e \left(\frac{h^2}{2\pi m_e kT}\right)^{3/2} = n_e\lambda_e^3,
\label{degPar}
\end{equation}
here  $\lambda_e=(h^2/2\pi m_e kT)^{1/2}$, i.e., the thermal de Broglie wavelength.
It should be noted also that
the effect of electron degeneracy manifests itself in EOS
via two channels: (i)~the degeneracy of free (unbound) electrons, which
is controlled by the parameter $\xi$, and (ii)~an indirect effect due to
the strong short-range repulsion of electrons localized within bound
complexes (atoms, molecules, atomic and molecular ions
etc).
Consequently, the resulting thermodynamics of
quasi-isentropically compressed gas is controlled by the
competition of the two strong effects: the nonideality because of  attraction
due to the average Coulomb interaction and the repulsion due to the electron
degeneracy.
As a result, one can ensure that at certain stages of
the steel shell --- pusher compression the thermodynamic trajectory of the
compressed gas (which mimics the ``collapse'' of a supernova core)
enters the region of the soft EOS corresponding to a combined
dissociation-ionization-driven (``plasma'') phase transition at
$P\sim 1-2$~Mbar with the density jump $\sim 15\,-\,20$\%.
This phase transition was discovered in explosive
experiments about a decade ago \cite{fortov:07} and has been many times
confirmed \cite{mochalov:10, mochalov:15, mochalov:17, mochalov:18}.
It should be stressed that some first indications on the possible existence
of another ionization-driven phase transition with a very high value of
the density jump are obtained at pressure $P\sim
50\, - \, 100$~Mbar in the latest high-explosive experiments \cite{mochalov:18}, see details
in Sec.~\ref{expData}. During crossing of the two-phase region(s) of
those phase transition(s) the EOS of the compressed deuterium
becomes very soft, so that the rate of the shell braking by the
pressure of the compressed gas goes down significantly (this mimics
the ``collapse'').
The shell is still compressible (driven by products of the
high explosion from outside) and is accelerating.
At the moment when all of the
deuterium goes into a close-packed phase of a strongly nonideal
($\Gamma \gg 1$) and strongly degenerate ($\xi\gg 1$) plasma, the rigidity
and resistance of deuterium plasma, hence, the rate of the shell braking,
increase sharply.
This causes a ``bounce'' of the falling matter, and a shock
at the bounce will run out into the shell.


\subsection{Two-cascade devices}

\begin{figure}[H]
\centering
\includegraphics[width=0.66\textwidth]{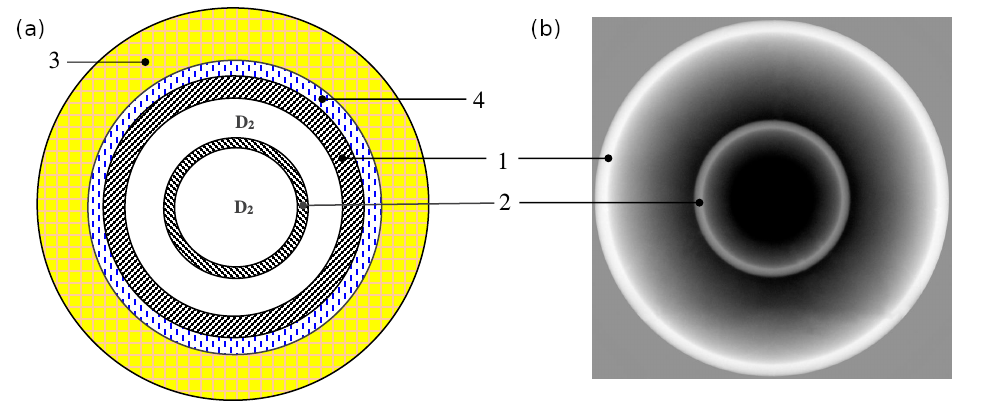}
\includegraphics[width=0.33\textwidth]{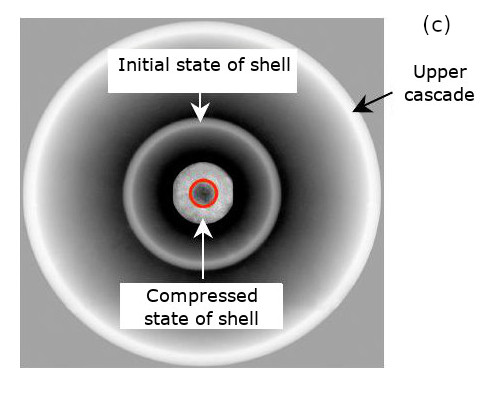}
\vspace{-0.5cm}
\caption{ (a) A schematic design of a two-cascade spherical experimental device: 1, shell 1 (Fe1);
2, shell 2 (Fe2); 3, explosive, 4 -- plexiglass;
(b) x-Ray patterns (roentgenograms) of the shells in the initial state.
(c) Initial and final states of the shell shown in one shot.
}
\label{fig:S2}
\end{figure}

A new type of experimental spherical device with separated cavities was recently
proposed to study the properties of plasma at high compression ratios.
This so-called two-cascade device has been developed and is now being used to
increase the compression ratio of plasma.
The device is schematically shown in Figs.~\ref{fig:SphereMochalov} and \ref{fig:S2}.
The compression of the gas and
plasma in such a construction is achived by the action of steel spherical shells~(1) and (2), see Fig.~\ref{fig:S2}.
The shells are accelerated to
the center of symmetry of the device by the  explosion of a powerful condensed
explosive~(3), made on
the basis of octogen, and by a system of shock waves reverberating in the cavity of the shell.
The inner cavity
of the shell~(2) is protected from the direct action of the explosive layer by the softening layer from the
test gas, which largely eliminates the ejection of metal particles into the internal plasma cavity. To further
reduce perturbations from the initiation system, a plexiglass gasket~(4) is used between the explosive unit~(3)
and the outer shell~(1).

\begin{figure}[H]
\centering
\vspace{-0.5cm}
\includegraphics[width=0.85\textwidth]{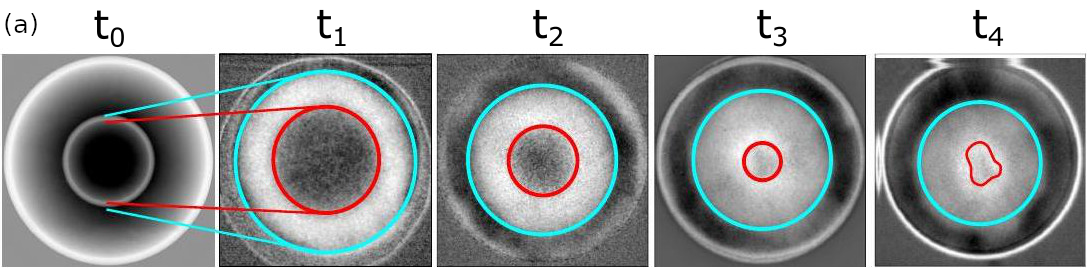}
\includegraphics[width=0.6\textwidth]{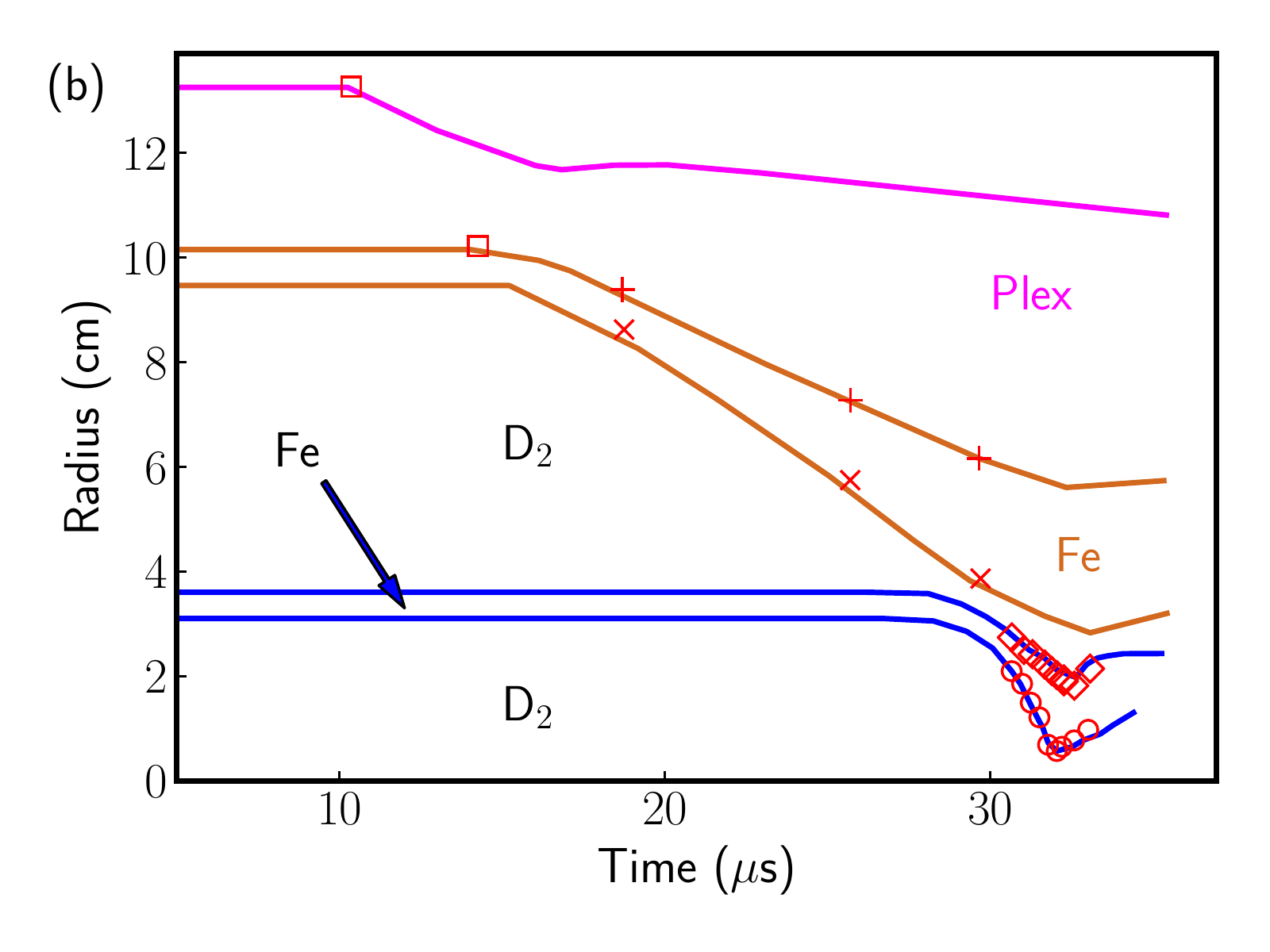}
\vspace{-0.5cm}
\caption{$R(t)$ diagrams for shells in the experimental device filled with deuterium
which produced the record high compression.
\newline
(a) x-Ray images of the shell 2 (see Fig.~\ref{fig:S2})
as a function of time, where $t_0$ is the initial state,
$t_1$ and $t_2$ are the compression phases, $t_3$ is the moment of maximum compression (``bounce'')
and $t_4$ is the expansion phase; light and red circles are the outer and inner boundaries of the shell 2,
respectively.
(b) experimental data and calculated $R(t)$ diagrams, symbol \Square\ denotes the results
of electrocontact technique, symbols $+$ and $\times$ are used for
x-ray data from the
model experiment, and symbols $\Diamond$ and \Circle -- for the data of the basic experiment.
}
\label{fig:RtDeuterium}
\end{figure}

Using devices of this type, the compressibility of deuterium plasma in the pressure region up to
$P \approx 5500$~GPa was investigated in Ref.~\cite{mochalov:17}.
At a ratio of the initial gas pressures in the inner and outer cavities $\approx 1 : 9$ with such a
construction, a helium plasma was compressed in Ref.~\cite{mochalov:15}
by a factor of 600 with pressure
$P \approx 3000$~GPa.
A spherical device similar to that described in Ref.~\cite{mochalov:17} 
was recently used in the experiment on the compression of deuterium plasma with
pressure of $10^4$~GPa~\cite{mochalov:18} (those new experiments were proposed
and discussed  in Ref.~\cite{S100}).
To compress the deuterium plasma to these high values of pressure, a significant amount of explosives is required: $m \approx 85$~kg TNT.
A number of x-ray images (roentgenograms)
obtained in the new experiment~\cite{mochalov:18} is shown in
Fig.~\ref{fig:RtDeuterium}.

To describe the process of plasma compression
as a function of time, one needs to carry out reliable numerical simulations.
The simulations must cover the propagation of shocks through the elements of the
structure and the motion of the shells in the initial phase (when the effect of the
gas is practically absent).
To select the gas-dynamic codes for those simulations,
a preliminary model gas-dynamic experiment was performed with a
hemispherical block simulating the geometry of the structure and the technology of the
full experiment.
In the preliminary experiment, the propagation of shocks through the plexiglass
(the motion of the shells in the initial phase) was recorded, and the velocity of the inner boundary
of the shell 1 in Fig.~\ref{fig:RtDeuterium} was measured.

The experimental data for the two-cascade device are shown in
Fig.~\ref{fig:RtDeuterium}(b)
together with the results of the gas-dynamic calculations.
From the graph in Fig.~\ref{fig:RtDeuterium}(b), it can be seen that the gasdynamic calculation satisfactorily
describes  the control points of the shock motion: the data from
the electroncontact technique in the plexiglass shell (symbol \Square), the
dynamics of the outer shell
boundaries by x--ray images (symbols $+$ and $\times$) and the data of the basic
experiment ($\Diamond$ and $\Circle$).
As follows from the
analysis of the calculations performed, at the moment of the maximum compression in
the deuterium plasma, the pressure has reached $P = (11400\pm 2000)$~GPa and the temperature
$T = 36\,500$~K with the measured density of the compressed plasma $\rho = 10.0 \pm 1.3$~g/cm$^3$
and the compression ratio $\sigma = \rho / \rho_0 = 300$.
The calculated value of the mean density $\rho_{\rm calc}= 11.1$~g/cm$^3$ agrees with the experimentally
measured value within 11\%.
The criterion for the validity of the maximum pressure value obtained in this way
is a good agreement between the
experimental data and the calculated $R(t)$ diagram.


Our goal in the current paper is to point out from those laboratory experiments the physics
that resembles supernova collapse.
This is allowed due to the
unique technique of generation and diagnostics  of matter in the extreme
states developed recently which allows matter compression by the energy
of high explosive charges (see, e.g., Ref.~\cite{Fortov:09} for a review).

As already mentioned, record levels of pressure in large volume $\sim50$~Mbar~\cite{mochalov:17} and a record compression ratio
($\rho/\rho_0 \sim 600$~\cite{mochalov:15}) have been achieved.
Pressures $P \sim 100$~Mbar (in large volumes)
have been achieved very recently~\cite{mochalov:18}, see Table.1.



\subsection{Details of diagnostics used in the experiment}

Very accurate simultaneity of initiation of the spherical and cylindrical
high-explosive charges,
achieved thanks to years of elaboration on this technique,
provides a high degree of symmetry of the collapsing steel shell, which
allows us to achieve a high degree of spherical symmetry of the  material motion
until the final moment of maximum compression of the
sample (when density grows by a factor of hundreds).
The symmetry of the driving shell is preserved until it stops before
the ensuing ``bounce''.

Extremely precise control of the parameters of the initial state is provided as well as control
by X-ray patterns of the position  of the moving (collapsing) shell,  and parameters of the compressible
material. The average density of the material is directly measured in the experiment from the position
of the shell along  its way until the time it stops just before its ``bounce''.

The pressure of compressed plasma is extracted from sophisticated hydrodynamic
calculations of the whole structure dynamics: the collapsing steel shell and the matter
compressed in the shock-wave reverberations.
The codes used in such simulations describe in needed detail the  multicomponent dynamics of
arbitrary matter phases: gases, solids, plasma, etc.
The accuracy of the description was tested many
times by comparison with experiments of similar type.
The codes rely on sophisticated equations of state described below in
Sec.~\ref{thermodynamics}.

The quasi-isentropic compression of the target gas is achieved by a sequence of steps.
The process starts by hitting the sample with an explosion-accelerated collapsing steel shell.
The first shock  converges toward the center and reflects there.
After the reflection, a much weaker diverging shock appears, and then this shock
is reflected by the steel shell, which is continuing to collapse.
Then, again, even weaker shock converging toward the center propagates, etc.,
up to the moment of the final stopping of the shell.

Numerous calculations do show that the bulk of entropy growth occurs in the first shock wave.
Further compression in a series of increasingly weak reverberating shocks can be
treated as isentropic in good approximation.
Hence, it was suggested to refer to this stage as
\textit{quasi-isentropic compression}~\cite{grigor:72,grigor:75,Fortov2016b}.

Table~\ref{Tab2} shows the main parameters achieved in various runs of
high-explosive experiments.
The most important quantity for comparison with core-collapsing supernovae
is the specific entropy per baryon (the last column).
It is close indeed to the predictions of the core-collapse simulations, see,
e.g. Fig.~\ref{fig:entropy}.

To detect the position of the shells that compress the tested material,
iron-free pulse betatrons (BIMs) are
widely used in devices with large metal masses and high explosives~\cite{Pavlovskii1965,Egorov2011}.
The average
density of the compressed material is measured along the inner boundary of the
shell with the  plasma at the moment of its maximum compression (the ``bounce'' moment).
Because the mass of the compressed matter is preserved, its density for a spherical device is calculated from the following simple expression:
\begin{equation}
                \rho = \rho_0 (R_0 / R_{\min})^n ,
\label{eq:Rratio}
\end{equation}
where $\rho_0$ is the initial gas density, $R_0$ and $R_{\min}$ are the inner shell radius in the initial state and at the
moment of its ``bounce'', respectively; and $n = 2$ or $n=3$ for cylindrical or spherical geometry, respectively.

The scheme of the experiment on the modern xray radiographic complex is shown in
Fig.~\ref{fig:SphereMochalov}, cf. Ref.~\cite{mochalov:17}.
A shadow image of the boundaries of the inner shell compressing the gas under investigation was obtained by
simultaneously using bremsstrahlung of three powerful betatrons~(1) with an electron limiting energy of
$\approx 60$~MeV located at $45^\circ$ angles to each other in a protecting concrete structure~(2).
A feature of the radiographic
complex is the possibility of each radiator to operate in a three-pulse mode, which allows one to register up
to nine phases of the shell motion in one experiment and thus to trace the entire dynamics of the target compression. When studying the motion of the shell for each betatron, an individual
optoelectronic detection system is used. 
The latter is activated
synchronously with the betatron pulses, which makes it possible to obtain three independent x-ray images.
To eliminate the effect of scattered radiation on highly sensitive recorders~(3), the size of the recording
field in each of the three projections is limited by the lead collimators~(4). To protect the betatrons~(1) and
optoelectronic x-ray detectors~(3), aluminum cones~(6) are used.
Single crystals of sodium iodide activated with tellurium NaI (Tl)
\O{}~150~mm ($\lambda_{\max} = 410$~nm, decay time 250~ns) and
lutetium silicate LSO \O{}~80~mm ($\lambda_{\max} = 420$~nm, decay time 50~ns) are used
as gamma converters in this system.

For additional technical details on diagnostics see Ref.~\cite{mochalov:18}.

\subsection{Experimental data}
\label{expData}

\begin{figure}
\centering
\includegraphics[width=0.6\textwidth]{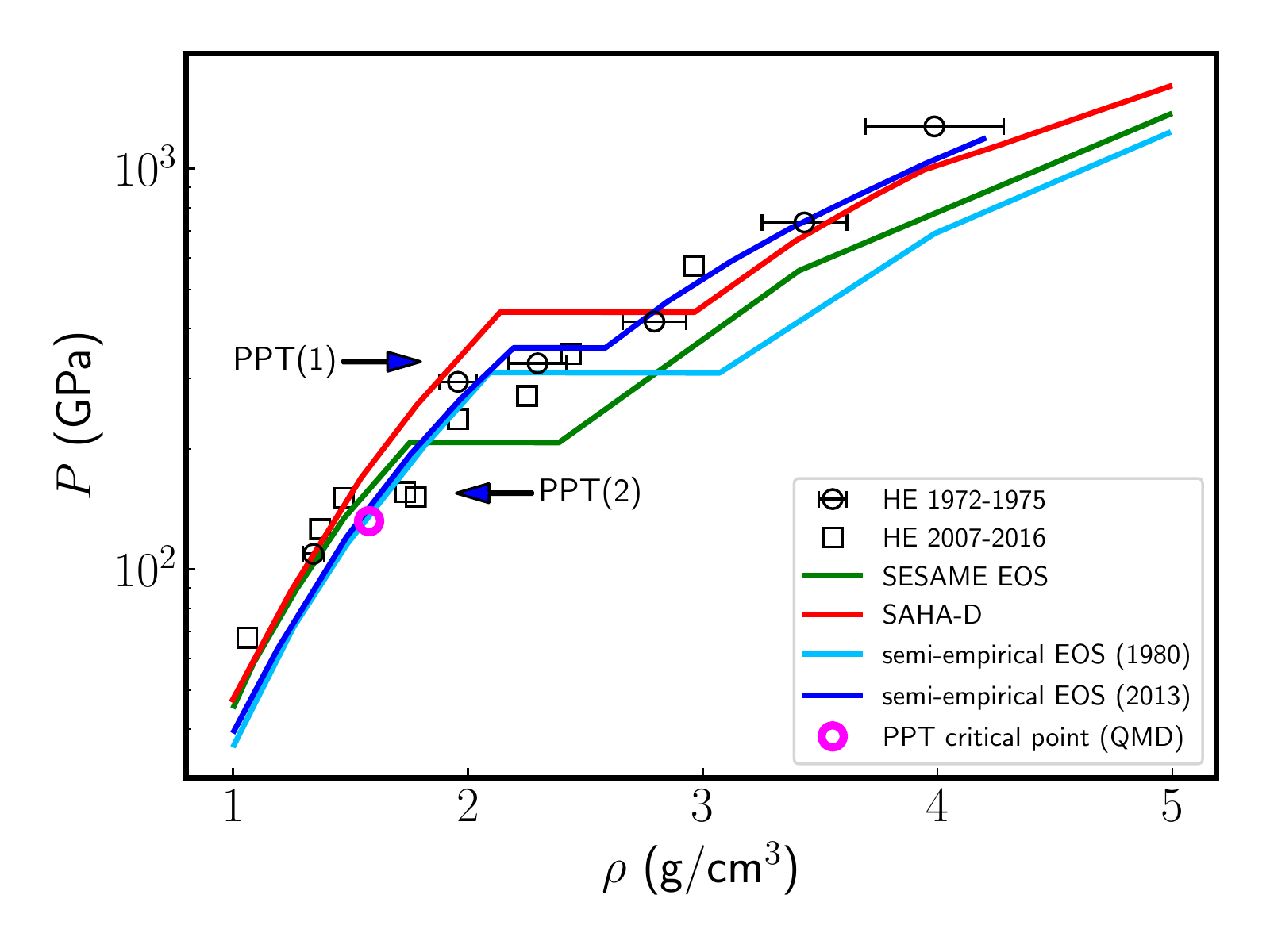} 
\vspace{-0.5cm}
\caption{Hypothetical phase transition in quasi-isentropic HE-driven compression experiments (1972--2017) and in ``cold'' (low entropy) theoretical models.
Experiment: hollow circles, quasi-isentropic VNIIEF 1972-1975~\cite{grigor:72, grigor:75}; rectangles -- quasi-isentropic compression
2007--2017 ~\cite{mochalov:10, mochalov:17}; arrows, phase-transition-like discontinuities (1) in Refs.~\cite{grigor:72, grigor:75} and (2)
in Refs.~\cite{mochalov:10, mochalov:17} supposed as ``plasma'' phase transitions (PPT).
Calculation results: green, isotherm $T = 0$~K from EOS SESAME~\cite{Kerley}; red line, isotherm $T = 300$~K from SAHA EOS ~\cite{2009JPhA...42u4007G} (``chemical picture''); light blue and blue, isotherm $T = 0$~K from two ``wide-range'' semiempirical
EOSes: Refs.~\cite{KopyshevKhrystalev1980} and~\cite{Urlin2013} correspondingly. Magenta circle, critical point of first-order liquid-liquid phase transition via {\it ab initio} QMD.}
\label{fig:EOSes}
\end{figure}

\begin{figure}
 \centering
\includegraphics[width=0.6\textwidth]{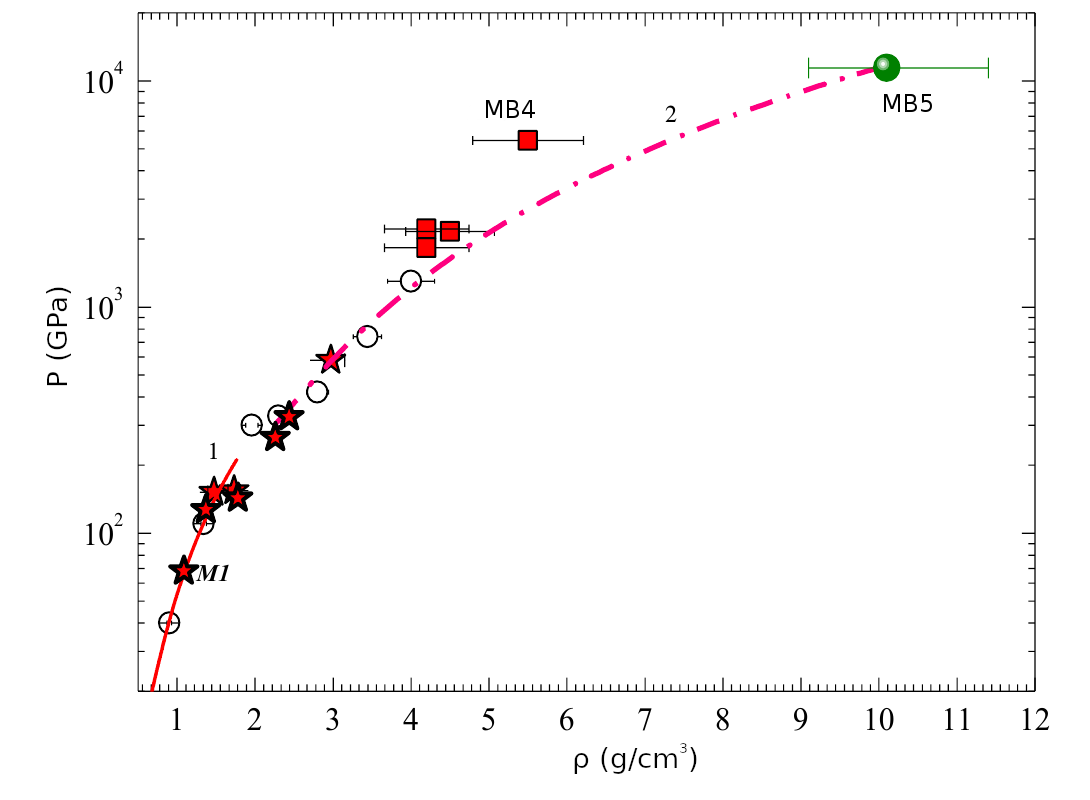}
 \vspace{-0.5cm}
\caption{Two isentropes: 1 is the isentropic curve from the point M1 with the use of
 VNIIEF EOS for  deuterium, and 2 is the isentrope based on SAHA-EOS  from Fig. 15 in
Ref.~\cite{mochalov:17}  above 150 GPa, which miraculously enters a new point MB5.}
\label{fig:isentrops}
\end{figure}

Figure~\ref{fig:EOSes} shows the experimental data that allow us to study equations of state and the dynamics of the multilayer system.
The results of this and other experiments are shown in Table~\ref{Tab1} and in Fig.~\ref{fig:isentrops}.
The data on the compressibility of deuterium plasma (obtained at pressure up to
$P \approx 5500$~GPa from Ref.~\cite{mochalov:17}) show
the density jump with $(\partial P / \partial \rho)_S \approx 0$ in the range
$\Delta\rho = 1.46 - 1.68$~g/cm$^3$  registered at
temperature $T\approx 3700$~K and pressure $P \approx 150$~GPa.
There is also a change in the slope of the derivative $dP/d\rho$
after the density jump (regardless of the magnitude $\Delta\rho$ of this jump).
These data are associated with a plasma phase transition~\cite{fortov:07, Fortov2016b, mochalov:17}.
The results of the current work and \cite{mochalov:18} indicate a new
change in the slope of the derivative $dP/d\rho$ at densities above
$\rho \sim 5$~g/cm$^3$ in a compressed plasma
of deuterium.
Those results hint to a new phase transition at $\rho\sim 5$~g/cm$^3$ that should be carefully studied in future work.
The development of
the necessary experimental devices for this study is not particularly difficult.

\section{Thermodynamics}
\label{thermodynamics}

The compressed matter is described by the SAHA-EOS model~\cite{2009JPhA...42u4007G} and the
corresponding SAHA-D code (see Ref.~\cite{mochalov:17} and references therein) in terms of the so-called
quasichemical representation (``chemical picture''), in other words, by the method of the ``free
energy minimization''.
Mutual transformations of components are described
according to the equations
of chemical and ionization equilibrium (such as the Saha equation) with corrections for nonideality.
A nontrivial point is that the hot, dense hydrogen plasma in experiments~\cite{grigor:72, grigor:75, mochalov:10,mochalov:18}
is strongly nonideal and strongly degenerate, i.e., the corresponding dimensionless parameters and corrections are not small.
This is so for the Coulomb parameter $\Gamma \gg 1$, introduced in (\ref{nonIdGamma}), 
and for the electron degeneracy
parameter $\xi \gg 1$ in (\ref{degPar}).

The most  ``rigorous'' approaches, \textit{ab initio}, or first principle EOS, based on the
so-called quantum Monte Carlo (QMC) and quantum molecular dynamics (QMD) ones, claim to reach
the status of a ``numerical experiment'', although it
remains to be seen whether they are accurate enough to describe real experiments.
They are very time-consuming and cumbersome, even in
the standard reference variant  Vienna Ab initio Simulation Package (VASP), a
plane wave density functional code for quantum molecular dynamics simulations, see,
e.g., Ref.~\cite{Chentsov:12}.
These methods  allow one to calculate directly only a part of the thermodynamic quantities,
the
``summation'' values of pressure $P (\rho, T)$ and internal energy $U (\rho, T)$.
They do not produce directly the truly full set of thermodynamic quantities (entropy, free
energy, and chemical potential).
The VASP technique also predicts the dissociative-plasma phase transition (see, e.g., Ref.~\cite{Lorenzen:10}).
For the discussion of the possible analogies of experiments with core-collapsing SNe, this is important.

HE-driven experimental results in Ref.~\cite{mochalov:17} and all the theoretical models (SAHA, DFT/MD, CH EOS, SESAME, Urlin~\cite{Urlin2013} )
account in someway for a phase transition of the 1st kind with a significant jump in density in the region of 1.5 -- 4 Mbar.

The basic analogy,
which can conceptually relate thermal and fluid dynamics in explosive experiments
with the processes occurring in supernovae of type II (core-collapsing supernovae, CCSN) may be formulated
as follows.

With the specially selected geometry of the experiment and the explosion intensity
one can ensure that at the final stages of compression
(``collapse'') of the steel shell the thermodynamic trajectory of the compressed deuterium would
start entering the two-phase domain of the phase transition.
When the central portion of the deuterium enters the two-phase zone, the EOS
of the deuterium plasma becomes soft and the rate of the shell braking stops growing.
The shell is still being compressed (``pushed'') by the products of explosion from outside, and it is accelerating.
At the moment when the whole deuterium gas enters the phase of a strongly degenerate plasma,
the stiffness of the deuterium plasma increases abruptly, and
the rate of the shell braking grows strongly.
A shock wave  will run out into the pushing shell, which will cause a rarefaction Riemann
wave on the outer boundary  and perhaps even a ``spallation'' of the surface.
There will be a ``rebound'' (or just bounce) of the shell.
In principle, this picture may  be observed in $\gamma$-rays probing the dynamics
of the whole process.

\section{Stellar collapse simulation}
\label{stCollSimul}

The death of massive stars is associated with one of the most striking events in the universe: a supernova explosion. The inner part of stellar core collapses to
nuclear density, where the nuclear EOS stiffens due to the repulsive
core of the nuclear force~\cite{bethe:90}.
Collapse is halted abruptly
on a millisecond timescale. The core of a newly born protoneutron star
overshoots its new equilibrium and then bounces back (``core bounce'')
into the still-infalling outer core, creating a shock wave.  This
shock first moves out dynamically but quickly loses energy by work
done to dissociate the infalling iron-group nuclei into neutrons, protons,
and $\alpha$ particles and also by the copious emission of neutrinos.

Many simulations of core-collapsing supernovae show that bounce pushes the shock wave,
but later it \emph{stalls}. Revival of the shock is needed for a successful explosion and it is
the most important problem in SNe that should be solved in the future. The physics of shock revival
is not reflected in experiments that we describe, but a study of the bounce with a laboratory
tool that we propose is essential, as this stage forms initial conditions for the shock stalling and
revival stages.

For a successful explosion to occur, the \emph{supernova mechanism}
must revive the shock within 1~s. 
Otherwise,
the steady accretion stream of outer core and shell material will push
the protoneutron star over its maximum mass (set by the nuclear EOS) and
black-hole
formation results (see, e.g., Ref.~\cite{oconnor:11}).
The primary candidate
mechanism for driving typical CCSN explosions is the \emph{neutrino
  mechanism}~\cite{colgate:60,arnett:66,nadyozhin:78,janka:12a}.
Neutrinos dominate CCSN energetics. The essence of the neutrino
mechanism is that a fraction ($\sim$$10\%$) of the outgoing $\nu_e +
\bar{\nu}_e$ luminosity is deposited in a heating region behind the front of the
stalled shock. This offsets the balance between the accretion ram pressure
and the total pressure behind the shock, eventually leading to a runaway
explosion~\cite{bethewilson:85}. 
However, the neutrino mechanism
\emph{fails} to explode ordinary massive stars in spherical symmetry (1D).
Extensive work~\cite{buras:06b,ott:08,mueller:12b,bruenn:16,lentz:15,couch:15a}
in axisymmetry (2D) and in 3D has shown that multidimensional (multi-D) fluid dynamics
may play a crucial role in the explosion mechanism.

The dynamics of the explosive experiment described above is similar in some aspects to the
process of collapse during supernova explosions. Here we present the results of our modeling of
a ``standard'' collapse which we will use for the illustration of our base analogy. We start from
a $2 M_\odot$ stellar iron core at the verge of its dynamical stability (i.e. the average adiabatic
index $\langle\gamma\rangle$ is slightly less than $4/3$). The core is divided into 1000
nonequally spaced Lagrangean mass zones. The description of hydrodynamic equations solver used
in this paper is given in Ref.~\cite{Yudin2009}. It is based on a number of routines, first
developed by D.~Nadyozhin and extensively used in various astrophysical applications, ranging
from core--collapse simulations \cite{IvanImshNad1969,Nadyozhin1977,Nadyozhin1977II} to low--mass
neutron star explosion processes \cite{Blinnikov1990}. After some additional modifications, this
solver now is a 1D, Newtonian, fully implicit Lagrangean FORTRAN code. It uses an artificial
viscosity
algorithm in a shock-capturing scheme when the shock is ``smeared'' onto three  mesh cells.
This simple approach appears, nevertheless, to be quite adequate when compared to much more
sophisticated methods, see, e.g., Ref.~\cite{Tolstov2013MNRAS}.

The matter at the subnuclear domain is assumed to be under nuclear statistical
equilibrium conditions, and the equation of state is taken according to
Ref.~\cite{NadYud2004}.
For the electron--positron plasma with the blackbody equilibrium radiation EOS part we use the code EPEOS~\cite{BlinDunNad1996}.
In the high-density domain, the effects of nonideality are included according to the excluded volume approximation \cite{Yudin2011}.
For uniform nuclear matter, formed at densities $\rho\geq 10^{14}~\mbox{g}/\mbox{ccm}$, we use Lattimer--Swesty type EOS parametrization~\cite{LattSwest1991}.

The most important part of the supernova simulation procedure, the neutrino transport scheme, is
divided into two parts: for the inner opaque stellar core domain we use neutrino heat
conduction (NHC) theory, first developed in Ref.~\cite{ImshNad1972} with additional scattering
effects~\cite{YudNad2008} included. For the outer semiopaque and transparent domain we use the
scheme proposed in Ref.~\cite{NadOtr1980} with a few modifications~\cite{Yudin2009}, which
ensure the smooth transition to the diffusion (NHC) limit.

\begin{figure}[htb]
\centering
\vspace{-0.2cm}
\includegraphics[width=0.5\textwidth]{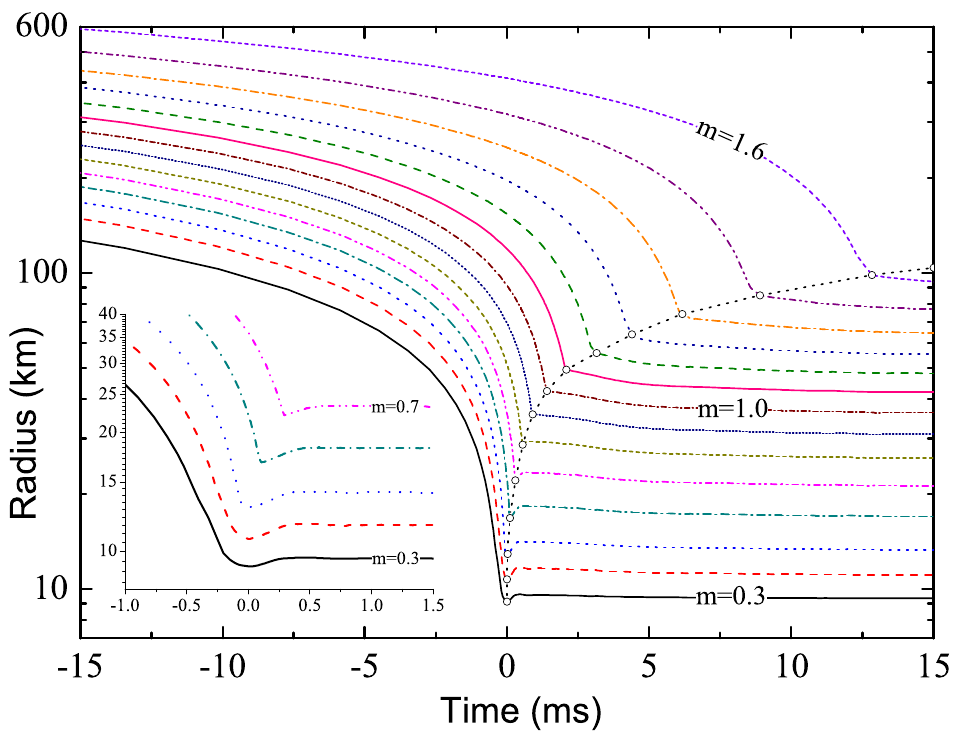}
\vspace{-0.2cm}
\caption{Evolution of radius coordinates during collapse and bounce for a
protoneutron star at several
fixed Lagrangean masses $m=0.3 \div 1.6~M_\odot$ (with $0.1~M_\odot$ step).
A dashed line, which connects
empty circles, shows the position of the shock. The lower-left corner of figure contains the zoomed-in part of the main image for the reduced time interval around the bounce and $m=0.3 \div 0.7~M_\odot$.}
\label{fig:SNbounce}
\end{figure}

Using our one--dimensional code,
described above,
we have obtained the trajectories of matter
inside the collapsing star; see Fig.~\ref{fig:SNbounce}. The dynamics here is shown
with so-called mass coordinates $m$, i.e., the mass enclosed by the radius $r$ (the
relation to ordinary coordinates is simple: $dm/dr=4\pi r^2\rho$, $\rho$ is the matter density).
We show the evolution of fixed Lagrangean masses $m=0.3 \div 1.6~M_\odot$ (with $0.1~M_\odot$ step) in the time interval
$-15\leq t\leq 15$~ms around the bounce (zero time).
The dashed line, connecting the empty circles, shows the position of the shock.
Behind the shock, the matter is only slightly compressed, forming a
quasiequilibrium configuration: a hot newborn neutron star.

The lower--left corner of the figure contains the zoomed-in part of the main image for a reduced time
interval around the bounce and $m=0.3 \div 0.7~M_\odot$.
Curves here are divided into two
types of behavior: for $m\leq 0.5$ they are smooth, and the matter is slightly overcompressed
and comes to a new equilibrium state. For $m\geq 0.6$ we see a ``kink''.
Such a difference signifies the appearance of the shock wave somewhere between $m = 0.5$
and $m = 0.6$ (see also discussion of Fig.~\ref{fig:entropy}).
For this and higher values of $m$ the matter falls until it meets the shock front and is
accelerated sharply: this leads to the formation of the ``kink'' in the enlarged pattern of the flow.

All of this is very similar to Fig.~\ref{fig:RtDeuterium} where snapshots of the record
high compression of the deuterium gas is reproduced. The comparison of this figures illustrate
our base idea: the qualitative similarity between these two, in principle very different, processes.

\begin{figure}[htb]
\centering
\vspace{-0.5cm}
\includegraphics[width=0.8\textwidth]{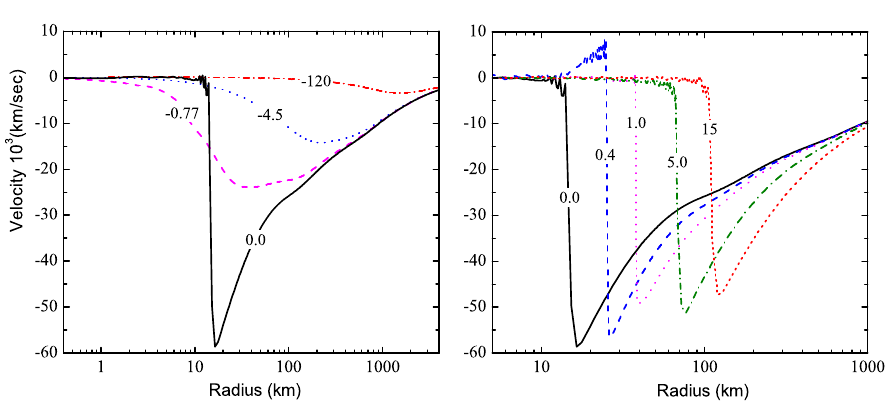}
\vspace{-0.5cm}
\caption{Velocity of matter as a function of $r$. Zero time is the moment
of bounce. Left and right panels show $v(r)$ for different moments before and after the bounce
correspondingly. The value of $t$ (in ms) is shown on each curve.
}
\label{fig:velocity}
\end{figure}

To show the process of collapse from another point of view, we plot in Fig.~\ref{fig:velocity}
the velocity of matter $v$ as a function of Eulerian coordinate $r$ for different moments of time
before (left panel) and after (right) the bounce. The moment $t=0$ corresponds to the bounce itself, and time
(in ms) is shown by numbers on each curve.
The shock wave is formed at 15~km approximately and starts
to propagate outwards.
At $t=0.4$~ms after bounce the velocity behind the shock is positive, but
soon the shock is converted into an accreting, although still expanding, one. By the moment
$t=1$~ms it moves to $r\approx 40$~km, then expands to $r\approx 70$~km at $t=5$~ms and at
the last  moment shown, $t=15$~ms, the shock is situated at $r=110$~km approximately.
Thus it decelerates and later this
outward moving shock wave stalls and is transformed into the standing accretion shock.
Naturally, the only mechanism of bounce is not sufficient for the explosion to occur.
Additional physics is required to revive this shock wave: neutrino contribution,
multidimensional effects in the flow, or others. This problem is actively discussed in the
literature~\cite{colgate:60,arnett:66,nadyozhin:78,janka:12a, buras:06b,ott:08,mueller:12b,bruenn:16,lentz:15,couch:15a}.
But, as mentioned above, the physics of the stalled shock
revival is not reflected in the experiments described in the current paper.

\begin{figure}[H]
\centering
\vspace{-0.5cm}
\includegraphics[width=0.8\textwidth]{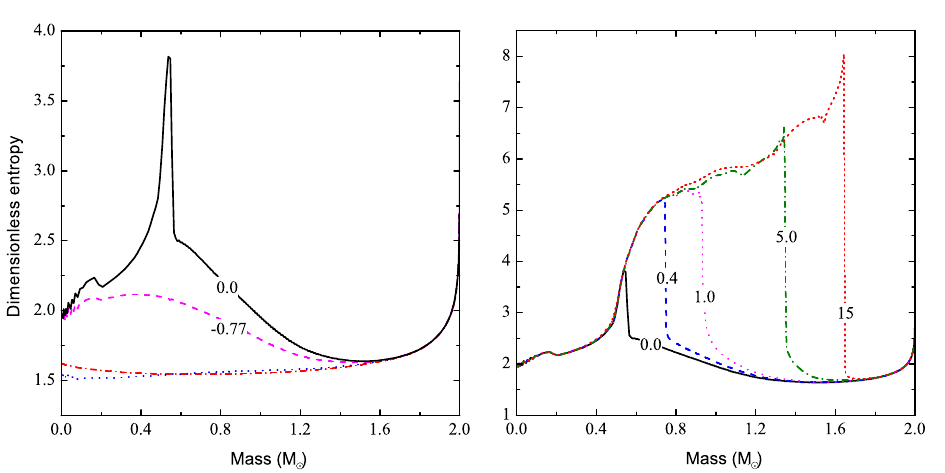}
\vspace{-0.5cm}
\caption{Entropy in the collapsing core.  Left panel shows entropy profiles as a function of Lagrangean mass
$m$ before the bounce and the right panel after. The moments of time are the same, as in Fig.~\ref{fig:velocity}.
}
\label{fig:entropy}
\end{figure}

Another important similarity between explosive experiments and core-collapse process is the low-entropy matter conditions.
Table~\ref{Tab2} (last column) shows the dimensionless entropy of deuterium plasma at the time of
maximum compression in our \emph{quasi-isentropic} experiments.
Now we can compare these data to the entropy values, reached during supernova explosion.

Figure~\ref{fig:entropy} shows dimensionless entropy per baryon in the collapsing stellar core for the same collapse model as in Figs.~\ref{fig:SNbounce} and \ref{fig:velocity}.
Left and right panels show the entropy profiles as a function of Lagrangean mass before and after the bounce respectively.
The moments of time are the same, as in Fig.~\ref{fig:velocity}.
One can see that before the bounce the entropy grows moderately until the shock   is formed
at $m\approx 0.55~M_\odot$ (see the discussion of the Fig.~\ref{fig:SNbounce} above).
Shock starts to propagate through the falling matter and heats it up.
But even behind the shock the values of entropy of matter are close by the order of magnitude to
the experimental results from Table~\ref{Tab2}.
Thus we can conclude that the low-entropy condition
is really another important similarity which connects stellar physics and our terrestrial
experiments.

\section{Summary and discussion}
\label{sec:summary}

The experiments presented in the paper reveal the analogy between explosive experiments for high compression of matter
and the dynamics of supernova collapses.

Let us point out which features of stellar core-collapse can be modeled by the experiments
of the same class as described in this paper.

\begin{itemize}
\item The analogy is based on the similarity of the dynamics: in both systems
a close-to-spherical compression occurs, the central region equation of state stiffens, and this leads to the bounce.
\item The important feature of the experiments is that all dynamics happens at low entropy: the
entropy value shows the relative contribution to the EOS of the cold part of matter, so the role of EOS changing
in the bounce process in laboratory reveals that in supernovae.
\item
The experimentally observed phase transition
of the first kind and softening of EOS may be relevant
for understanding the exotic mechanisms of supernova explosions, where a phase transition
may occur.
The phase transitions which are plausible in SN dynamics~\cite{2009PhRvL.102h1101S} may help to
produce a successful supernova~\cite{2018NatAs...2..980F} in the process of evolution of
massive progenitors.
\item The development of 3D hydrodynamic instabilities may be studied in future experiments
with controlled initial asymmetric perturbations and/or addition of magnetic fields.
\item Finally, the spallation of the surface of compressed samples after the bounce is similar to another phenomenon already observed in
supernovae, namely, the so-called shock breakout~\cite{schaw:08,gezari:15,garnavich:16}.
\end{itemize}

The reader should be aware that we are able to model only a limited subset of phenomena
occurring in a real stellar collapse.
The following features are not reproduced in HE experiments:

\begin{itemize}
 \item Neutrino emission.
 \item Dissociation of nuclei by shock waves.
 \item Shock wave stalling in the accretion flow.
 \item Gravity and General Relativity (GR) effects.
\end{itemize}

The goal of our paper is to emphasize the analogy and to present such experiments as a new platform for laboratory astrophysics.
Consequently, the experiments of this kind expand the possibilities and the list of existing
laboratory astrophysics platforms (for a review see~\cite{2006RvMP...78..755R, Fortov2016}), among which we could highlight  the following.
\begin{itemize}
\item High-compression experiments (see, e.g., Refs.~\cite{azechi:91, azechi:2016}) that study equations of state in conditions similar to interiors of giant planets.
\item Laboratory and numerical experiments on hydrodynamic instabilities and opacity measurements in relation to stellar
and supernova physics~\cite{2000PhPl....7.1662V,2003PhPl...10.1883Z,2006RvMP...78..755R}.
\item Experiments on collisionless physics of gaseous supernova remnants (e.g. collisionless shock waves) \cite{2017APS..DPPCO5002P}.
\end{itemize}

It should be pointed out that previous work on laboratory astrophysics of core-collapsing
supernovae was concentrated more on problems of Rayleigh-Taylor and Richtmyer-Meshkov
instabilities in the supernova shocks \cite{2000PhPl....7.1662V,2003PhPl...10.1883Z,2006RvMP...78..755R}.
More recent work is done on the simulations growth of instabilities in the collapse phase,
see Ref.~\cite{2014JCoPh.275..154J} and references and citations therein.

Our paper presents  a new record of compression in laboratory high explosive experiments.
The pressure obtained in experiments of another type,  namely, laser facilities, e.g. NIF~\cite{hurricane:16},
may be higher, approaching $P\sim 100$~Gbar, but at significantly
smaller spatial scales and short lifetimes. The explosive experiment described by us reaches maximum
pressure at $l_{\rm expl}\sim 1$~cm scale and $t_{\rm expl}\sim 1$~$\mu$s
(cf. Fig.~\ref{fig:RtDeuterium}),
while for
laser facilities these numbers are $l_{\rm las}\sim 10 \mu$m scale and $t_{\rm expl}\sim 0.1-1$~ns.
The latter significantly complicates the diagnostics. The larger scale of
explosion-driven compression allows one to get much more details~\cite{1998PhT....51h..26H,
Drake, Fortov2016b, Fortov2016} of the hydrodynamic flows and shocks.

The goal of our paper is not to give an exhaustive review of the current status of core-collapse
simulations with successful and unsuccessful SN explosions (see
Refs.~\cite{buras:06b,ott:08,mueller:12b,bruenn:16,lentz:15,couch:15a} and the papers citing them).

One may ask, Are there any scenarios in which one might expect the shock generated
in the HE experiment to ``fail'' and would there be any secondary
indications of this failure on the plasma properties?
In fact, our experiments allow us to model only a limited number of properties.
The shocks propagating in the compressed material are not stalled by the energy losses due to
dissociation of heavy nuclei and huge neutrino emission from the downstream region, as is
envisaged in supernovae.
Those features are not possible to model in the HE experiments.
However, even the narrow range of properties in the experiment is useful for understanding the
flows in nonideal plasma which are initially
spherically symmetric with a high degree of accuracy during the compression phase and later
develop 3D asymmetries at the expansion stage.

There have been suggestions that even failed supernovae can give
rise to the secondary indications of the collapse: weak shocks and outbursts
(e.g., Refs.~\cite{1980Ap&SS..69..115N,2013ApJ...769..109L}).
These conclusions  are relevant for
more realistic situations when neutrinos take away a lot of energy from the collapsing core.
Nevertheless, those neutrino losses occur on a long diffusion timescale
(seconds in the star core), while the bounce of the stellar core goes on a much shorter
timescale (milliseconds) after neutrinos are trapped.
Thus, the pattern of hydro flows is very similar in our experiments and in the simulations of the stellar core bounce.

The recent progress in supernova theory shows, that even collapses which do not lead to a
prompt supernova explosion may have bright manifestations due to a fallback
energy release, see Refs.~\cite{2018MNRAS.475L..11M,2018MNRAS.477.1225C}.
Reference~\cite{2018MNRAS.477.1225C} not only develops a physical model of mass ejection
in failed supernovae but also advances a self-similar solution applicable for those events.
In future work, self-similar solutions of this type may be tried also for the description of
hydrodynamic flows in experiments discussed here (cf. the results on self-similar volume
compression in laser fusion conditions~\cite{1998JPlPh..60..743H}). 

\section{Conclusions}
\label{sec:conclusions}

The similarity in dynamics seen in Figs.~\ref{fig:RtDeuterium} and \ref{fig:SNbounce} is
a basic analogy, which can conceptually relate thermal and fluid dynamics in explosive
experiments with the processes occurring in type Ib/c and type II supernovae
(CCSN) and permits us to use the explosive experiment as a
laboratory site to study hydrodynamics of the collapse and bounce.
Not only a new dimension for laboratory astrophysics is open by the new experiments
but also a new tool appears for validation of codes used in applied science and in astrophysics.

The simulations of explosive experiments with a hydrodynamic code that describes the
dynamics of the whole structure, the steel shell, the gas-target, and high explosive and
detonation products, agrees well with
the results of the experiment. 
An important part of the simulations are equations of state, that describe
all transitions that occur within the material being investigated.
As was emphasized above, these transitions are crucial for the proposed analogy between
supernovae and explosive experiment.
The equations of state used in experiment description are the realization of theoretical models~\cite{Fortov2016b, Fortov2016}, which is published elsewhere~\cite{mochalov:17}.
Another important feature of the experiment is the low entropy regime of compression, this
fact brings the experiments closer to the condition in real supernovae, where entropy of
the collapsing material is also quite low.

We conclude that such an experimental tool opens new horizons in laboratory
astrophysics. It allows one to study the process of collapse: the compressed matter
mimics this astrophysical phenomenon. High pressures together with low entropy lead to
degeneracy of plasma and, therefore, simulate the stiffness of the equation of state in
a real collapsing star, leading to the bounce of the shell. This process is an inevitable
part of collapse and its hydrodynamics can now be studied in laboratory. Full
control of initial conditions in laboratory gives a possibility to investigate the
role of additional effects on the hydrodynamics of collapse, e.g. perturbations that
violate the spherical symmetry of the system (this resembles asymmetries in initial star
configuration, like rotation, magnetic field,  etc.).

%
%
%


\acknowledgments
We are grateful to V.E.Fortov for the idea to use HE experiments in laboratory astrophysics
of bounce in core-collapsing supernovae and to anonymous referees for valuable comments.

 This research was partially supported by
 Russian Science Foundation Grant No. 16-12-10519 (S.~Glazyrin)
 and Grant No. 18-12-00522 (S.~Blinnikov).
%



\newpage

\begin{table}[H]
\caption{Parameters of deuterium plasma compressed by pressure
$P\approx 11400$~GPa (experimental values and simulation with VNIIEF EOS)}
\begin{center}
\begin{tabular}{|l|c|c|c|c|c|c|c|}
\hline
run &  $R_0$, mm & $R_{\min}$, mm &  $\rho_0$, g/cm$^3$ & $\rho_{\exp}$, g/cm$^3$ &  $P_{\rm calc}$, GPa  &  $\rho_{\rm calc}$, g/cm$^3$ &  $T_{\rm calc}$, kK \\
\hline
MB5 &  31 &  4.74 &0.0354&  $10.1^{+1.3}_{-0.9} $ &  $11400^{+2000}_{-2000}$ &  11.1 &  36.5 \\
\hline
\end{tabular}
\end{center}
\label{Tab1}
\end{table}


\begin{table}[H]
\caption{Deuterium plasma state at the time of maximum compression in quasi-isentropic compression experiments (Vserossiiskiy
Nauchno-Issledovatelskiy Institut Experimentalnoi Fisiki,
 VNIIEF, Sarov),
calculated using EOS SAHA. Given are values of pressure $P$, density $\rho$, temperature $T$,
degeneracy parameter for free electrons $\xi = n_e \lambda^3$ (\ref{degPar}), and specific entropy $S$ (per gram and per nucleon)
}
\begin{center}
\begin{tabular}{|l|c|c|c|c|c|c|}
\hline
run &  $P$, GPa  &  $\rho$, g/cm$^3$ &  $T$, kK &  $n_e \lambda^3$ & $S$, J/g\textperiodcentered K & $S$,
$1/k_{\rm B}$  \\
\hline
1 &  68  &  1.07 &  2.15 &  --   &  22.2 &  2.67 \\
\hline
2 &  127 &  1.35 &  2.49 &  8.04 &  23.5 &  2.83 \\
\hline
3 &  143 &  1.76 &  2.60 &  9.6  &  -- &  -- \\
\hline
4 &  265 &  2.2  &  4.52 &  60.5 &  29.4 &  3.53 \\
\hline
5 &  327 &  2.37 &  6.30 &  54.7 &  31.8 &  3.82 \\
\hline
6 &  583 &  2.91 &  6.85 &  66.3 &  30.9 &  3.72 \\
\hline
7 &  1830 &  4.2 &  19.71 &  36.0 &  37.6 &  4.52 \\
\hline
8 &  2215 &  4.2 &  31.50 &  21.4 &  41.8 &  5.03 \\
\hline
9 &  2160 &  4.5 &  21.15 &  35.9 &  37.7 &  4.53 \\
\hline
10 &  5450 &  5.5 &  69.16 &  11.4 &  46.8 &  5.63 \\
\hline
MB5 & $11400^{+2000}_{-2000}$ &  11.1 &  36.5 &  -- & 34.5 & 4.15 \\
\hline
NIF &  $1.4\cdot 10^7$ &  30 &  $5\cdot 10^4$ &  -- &  117 &  14.1 \\
\hline
\end{tabular}
\end{center}
\label{Tab2}
\end{table}

\end{document}